\documentclass[fleqn,oneside]{article}
\usepackage{espcrc2}

\IfFileExists{srcltx.sty}{\usepackage[active]{srcltx}}

\usepackage{graphicx}
\usepackage[figuresright]{rotating}

\newcommand{\AmS}{{\protect\the\textfont2
  A\kern-.1667em\lower.5ex\hbox{M}\kern-.125emS}}

\title{Ultrahigh-energy nuclei, photons, and magnetic fields}

\author{Alexander Kusenko\address{Department of Physics and Astronomy, University of California, Los
Angeles, CA 90095-1547, USA
}%
\address{IPMU, University of Tokyo, Kashiwa, Chiba 277-8568, Japan
}
        }
       
\begin{document}
\begin{abstract}
Combined recent data from cosmic-ray detectors and gamma-ray detectors have produced some surprising insights regarding the sources of ultrahigh-energy cosmic rays (UHECRs), magnetic fields inside and outside the Milky Way, and the universal photon backgrounds.  The energy-dependent composition of UHECRs implies a non-negligible contribution of sources located in the Milky Way, such as past gamma-ray bursts that took place in our Galaxy.  Extended halos of distant sources seen in the {\em Fermi} data imply that intergalactic magnetic fields have average strengths of the order of a femtogauss.  Such relatively  low magnetic fields imply that the protons from distant blazars generate a detectable flux of secondary gamma rays in their interactions with the photon background.  A comparison with 
the data shows an excellent agreement of the secondary photons with the spectra of distant blazars observed by atmospheric Cherenkov telescopes. 

\end{abstract}

\maketitle


\section{Introduction}

Several surprising insights have emerged from the recent data on ultrahigh-energy cosmic rays and very high energy gamma rays. 

Pierre Auger Observatory (PAO) has reported a steady increase of the mean nuclear mass with energy between 2~EeV and 35~EeV~\cite{Abraham:2009dsa,Abraham:2010yv}. This unexpected result is difficult to reconcile with the usually assumed extragalactic origin of UHECR.  Indeed, most candidate sources of UHECRs, such as active galactic nuclei (AGN), are characterized by a relatively high photon density in the region of acceleration, which implies that only a small fraction of nuclei can survive and come out of the source.  In addition, UHECR nuclei are prone to disintegration in photon backgrounds, which limits their propagation distances.   

However, it was recently pointed out that, if the sources are located inside the Milky Way Galaxy, diffusion in the galactic magnetic fields could produce the spectrum and composition pattern consistent with the PAO data~\cite{Calvez:2010uh}.  This brings in sharper focus the possibility that past GRBs in Milky Way Galaxy, which are expected to occur on the time scales shorter than the diffusion times of heavy nuclei, can be responsible for a large fraction of UHECR between 1 and 30~EeV.  
Furthermore, if the cosmic rays of 10$^{18}-10^{19}$~eV are nuclei produced in the Milky Way, the effects of diffusion in turbulent Galactic micro-Gauss magnetic fields can can also explain the lack of galactocentric anisotropy~\cite{Calvez:2010uh}.

The lack of absorption features in the gamma-ray spectra of distant blazars can be explained by the production of secondary photons by protons emitted by these blazars~\cite{Essey:2009zg,Essey:2009ju}.  This explanation holds as long as the intergalactic magnetic fields (IGMF) are not strong enough to deflect the protons significantly.  The recent measurement of IGMF using the {\em Fermi} data shows these fields have femtogauss strengths~\cite{Ando:2010rb}, which supports the aforementioned interpretation. 

\section{GRBs and unusual supernovae in the Milky Way}

There is a growing evidence that long GRBs are caused by a relatively rare type(s) of supernovae, while the short GRBs probably result from the coalescence of neutron stars with neutron stars or black holes. Compact star mergers undoubtedly take place in the Milky Way, and therefore short GRBs should occur in our Galaxy. 

Although there is some correlation of long GRBs with star-forming metal-poor galaxies~\cite{Fruchter:2006py}, many long GRBs are observed in high-metallicity galaxies as well~\cite{Savaglio:2006xe,CastroTirado:2007tn,Levesque:2010rn}, and therefore one expects that long GRBs should occur in the Milky Way.  Less powerful hypernovae, too weak to produce a GRB, but can still accelerate UHECR~\cite{Wang:2007ya}, with a substantial fraction of nuclei~\cite{Wang:2007xj,Murase:2008mr}.  

If the observed cosmic rays originate from past explosions in our own Galaxy, PAO results have a straightforward explanation~\cite{Calvez:2010uh}.  

GRBs have been proposed as the sources of extragalactic UHECR~\cite{Waxman:1995vg,Vietri:1995hs,Murase:2008mr}, and they have also been considered as possible Galactic sources~\cite{Dermer:2005uk,Biermann:2003bt,Biermann:2004hi}. It is believed that GRBs happen in the Milky Way at the rate of one per  $t_{GRB}\sim 10^{4}-10^{5}$ years \cite{Schmidt:1999iw,Frail:2001qp,Furlanetto:2002sb,Perna:2003bi}. Such events have been linked to the observations of positrons~\cite{Bertone:2004ek,Parizot:2004ph,Parizot:2004GRB,Ioka:2008cv,Calvez:2010fd}. 

If local sources, such as past GRBs in the Milky Way, produce a small fraction of heavy nuclei, the observed fraction of UHE nuclei is greatly amplified by diffusion. This  is because the galactic magnetic fields are strong enough to trap and contain nuclei but not protons with energies above EeV.  This observation leads to a simple explanation of the composition trend observed by PAO.

\section{Effects of diffusion}

As illustrated in Fig.~1, diffusion depends on rigidity, and, therefore, the observed composition can be altered by diffusion~\cite{Calvez:2010uh,Wick:2003ex}. Changes in composition due to a magnetic fields have  been discussed in connection with the spectral ``knee''~\cite{Wick:2003ex}, and also for a transient source of UHECR~\cite{Kotera:2009ms}.  The ``knee'' in the spectrum occurs at lower energies than those relevant PAO, and at higher energies the cosmic rays effectively probe the spectrum of magnetic fields on greater spatial scales, of the order of 0.1~kpc~\cite{Han:2004aa}.  
\begin{figure}
   \includegraphics[width=.45\textwidth]{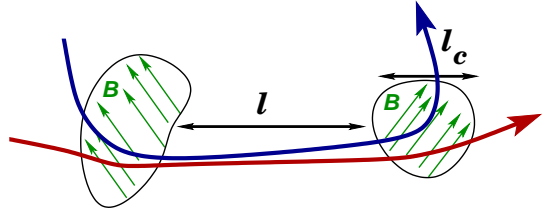}
  \caption{For each species, there is a critical energy $E_{0,i}$ for which the Larmor radius $R_i$ is equal to the magnetic coherence length $l_c$. For $E\ll E_{0,i}$, the mean free path of the diffusing particle is $l \sim l_0$, and $D_i(E)= l_c/3$.  For $E\gg E_{0,i}$, the particle is deflected only by a small angle $\theta\sim l_0/R_i$, and, after $k$ deflections, the mean deflection angle squared is $\bar{\theta^2}\sim k (l_0/R_i)^2$.  The corresponding diffusion coefficient is $D_i(E) \propto ( \frac{E}{E_{0,i}})^2 , \ {\rm for} \ E \gg  E_{0,i}$.}
\end{figure}

One can use a simple model~\cite{Calvez:2010uh} to show how diffusion affects the observed spectrum of the species ``$i$'' with different rigidities.  Let us suppose that all species are produced with the same spectrum $n_i^{(\rm src)}=n_0^{(\rm src)}\propto E^{-\gamma} $ at the source located in the center of Milky Way and examine the observed spectra altered by the energy dependent diffusion and by the  trapping in the Galactic fields.
 In diffusive approximation, the transport inside the Galaxy can be described by the equation:
\begin{eqnarray}
\frac{\partial n_i}{\partial t} - \vec{\nabla}(D_i \vec{\nabla}n_i)  +\frac{\partial}{\partial E}(b_i n_i )= \nonumber \\
 Q_i(E,\vec{r},t) +\sum_k\int P_{ik}(E,E') n_k(E') dE'.
\label{transport}
\end{eqnarray}
Here $D_i(E,\vec{r},t)=D_i(E)$ is the diffusion coefficient, which we will assume to be constant in space and time. The energy losses and all the interactions that change the particle energies
are given by $b_i(E)$ and the kernel in the collision integral $P_{ik}(E,E')$. For energies below GZK cutoff, one can neglect the energy losses on the diffusion time scales.

The diffusion coefficient $D(E)$ depends primarily on the structure of the magnetic fields in the Galaxy. Let us assume that the magnetic field structure is comprised of uniform randomly oriented domains of radius $l_0$ with a constant field $B$ in each domain. The density of such domains is $N\sim l_0^{-3}$. The Larmor radius depends on the particle energy $E$ and its electric charge $q_i=eZ_i$:
\begin{eqnarray}
R_i & =&   l_0 \left( \frac{E}{E_{0,i}} \right),\ {\rm where} \
E_{0,i}= E_0 \, Z_i, \\
E_0 & = &  10^{18} {\rm eV} \left(\frac{B}{3\times 10^{-6}\, {\rm G}}\right) \left(\frac{l_0}{0.3\, {\rm kpc}} \right)   . \label{E0i}
\end{eqnarray}

The spatial energy spectrum of random magnetic fields inferred from observations suggests that $B\sim 3\mu {\rm G} $ on the 0.3~kpc spatial scales, and that there is a significant change at $l=1/k\sim 0.1-0.5$~kpc~\cite{Han:2004aa}. This can be understood theoretically because the turbulent energy is injected into the interstellar medium by supernova explosions on the scales of order 0.1~kpc.  This energy is transferred to smaller scales by direct cascade, and to larger scales by inverse cascade of magnetic helicity.  Single-cell-size models favor $\sim 0.1$~kpc scales as well~\cite{Han:2004aa}.

As explained in the caption of Fig.~1, diffusion occurs in two different regimes depending on whether the Larmor radius is small or large in comparison with the correlation length.  As a result, the diffusion coefficient changes its behavior dramatically at $E=E_{0,i} $:

 \begin{equation}
 D_i(E) = \left \{
\begin{array}{ll}
D_0 \left ( \frac{E}{E_{0,i}} \right)^{\delta_1}, & E \le  E_{0,i}, \\
D_0 \left ( \frac{E}{E_{0,i}} \right)^{(2-\delta_2)},  & E > E_{0,i} .
\end{array} \right.
\end{equation}

Here the two parameters $0 \le \delta_{1,2} \le 0.5$ are different from zero if the magnetic domains are not of the same size. The exact values of 
these parameters depend on the power spectrum of turbulent magnetic fields. 

The approximate solution of the transport equation in our simple model yields 
\begin{equation}
n_i(E,r) = \frac{Q_0}{4\pi r\, D_i(E)} \left ( \frac{E_0}{E} \right )^\gamma.
\label{solution_GC}
\end{equation}
Since diffusion depends on rigidity, the composition becomes energy dependent.  Indeed, at critical energy $E_{0,i}$, which is different for each nucleus, the solution (\ref{solution_GC}) changes from $\propto E^{-\gamma} $ to $\propto E^{-\gamma-2} $ because of the change in $D_i(E)$, as discussed in the caption of Fig.~1.  Since the change occurs at a rigidity-dependent critical energy $E_{0,i}=e E_0 Z_i$, the larger nuclei lag behind the lighter nuclei in terms of the critical energy and the change in slope.  If protons dominate for $E<E_0$, their flux drops dramatically for $E>E_0$, and the heavier nuclei dominate the flux.  The higher $Z_i$, the higher is the energy at which the species experiences a drop in flux.

One can also understand the change in composition by considering the time of diffusion across the halo is $t_i \sim R^2/D_i$.  The longer the particle remains in the halo, the higher is the probability of its detection.   At higher energies, the magnetic field's ability to delay the passage of the particle diminishes, and the density of such particles drops precipitously for $E>E_{0,i}$.  Since $E_i$ is proportional to the  electric charge, the drop in the flux occurs at different energies for different species.

\begin{figure}
  \includegraphics[width=.35\textheight]{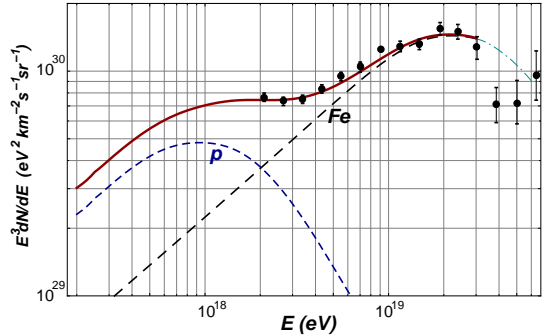}
  \caption{UHECR spectra according to the model of Ref.~\cite{Calvez:2010uh}, shown here for different values of magnetic fields (to illustrate the robustness of the model).
The magnetic field was assumed to be $\sim 10\mu$G, coherent over $l_0=100$~pc domains ({\em cf.} $B\sim 4\mu$G, $l_0=200$~pc in Ref.~\cite{Calvez:2010uh}). The power and the iron fraction were adjusted to fit the PAO data~\cite{Abraham:2009wk}. 
}
\end{figure}

The model of Ref.~\cite{Calvez:2010uh} gives a qualitative description of the data. To reproduce the data more accurately, it must be improved.  First, one should use a more realistic source population model.  Second, one should include the coherent component of the Galactic magnetic field.  Third, one should not assume that UHECR comprise only two types of particles, and one should include a realistic distribution of nuclei. Finally, one should include the extragalactic component of UHECR produced by distant sources, such as active galactic nuclei (AGN) and GRBs (outside the Milky Way). 

A recent realization that very high energy gamma rays observed by Cherenkov telescopes from distant blazars are likely to be secondary photons produced in cosmic ray interactions along the line of sight lends further support to the assumption that cosmic rays are copiously produced in AGN jets~\cite{Essey:2009zg,Essey:2009ju}. For energies $E>3\times 10^{19}$~eV, the energy losses due to photodisintegration, pion production, pair production and interactions with interstellar medium become important and must be included. The propagation distance in the Galaxy exceeds 10~Mpc, so that the Galactic component should exhibit an analog of GZK suppression in the spectrum.  The extragalactic propagation can also affect the composition around $10^{18}$~eV~\cite{Hill:1983mk}.

Galactocentric anisotropy for a source distribution that traces the stellar counts in the Milky Way is small~\cite{Calvez:2010uh}. Although the anisotropy in protons is large at high energies, their contribution to the total flux is small, so the total anisotropy was found to be $<10\%$, consistent with the observations.  The latest GRBs do not introduce a large degree of anisotropy, as it would be in the case of UHE protons, but they can create ``hot spots'' and clusters of events.

The model of Ref.~\cite{Calvez:2010uh} makes an interesting prediction for the highest-energy cosmic rays.  Just as the protons of the highest energies escape from our Galaxy, they should escape from the host galaxies of remote sources, such as AGN. Therefore, UHECR with $E>3\times 10^{19}$~eV should correlate with the extragalactic sources.  Moreover, these UHECR should be protons, not heavy nuclei, since the nuclei are trapped in the host galaxies.  If and when the data will allow one to determine composition on a case-by-case basis, one can separate $E>3\times 10^{19}$~eV events into protons and nuclei and observe that the protons correlate with the nearby AGN.  This prediction is one of the non-trivial tests of our model: at the highest energies the proton fraction should exist and should correlate with known astrophysical sources, such as AGN.  The microgauss magnetic fields in Milky Way cause relatively small deflections for the highest-energy protons.  As for the intergalactic magnetic fields, the first detection and a measurement of such fields, using the data from {\em Fermi Gamma-ray Space Telescope}, points to relatively weak, femtogauss field strengths~\cite{Ando:2010rb}, which should not affect the  protons significantly on their trajectories outside the clusters of galaxies.   

If local, Galactic GRBs are the sources of UHECRs, the energy output in cosmic rays should be of the order of $10^{46}$~erg per GRB.  This is a much lower value than what would be required of extragalactic GRBs to produce the same observable flux.  Indeed, in our model the local halo has a much higher density of UHECR than intergalactic space, and so the overall power per volume is much smaller.  The much higher energy output required from extragalactic GRBs~\cite{Waxman:1995vg,Vietri:1995hs,Murase:2008mr} in UHECR has been a long-standing problem.  The same issue does not arise in our case because it seems quite reasonable that a hypernova or some other unusual supernova explosion would generate $10^{46}$~erg of UHECR with energies above 10~EeV.  

\section{Intergalactic magnetic fields}

A recent analysis of the stacked images of 170 AGN observed by {\em Fermi Gamma-Ray Space Telescope } has made possible the first measurement of the intergalactic magnetic fields~\cite{Ando:2010rb}.  
The composite image of multiple AGN would be pointlike, broadened only by the instrumental effects, in the absence of IGMF.  However, the images obtained from {\em Fermi} data exhibit halos consistent with the deflections of electromagnetic cascades by IGMF.  Two alternative analyses, based on the pre-launch calibration and on the actual image of the Crab pulsar, yield consistent and statistically 
significant evidence that the halos are real, and that they cannot result from an instrumental effects.  Furthermore, the halo parameters depend on the redshifts of the sources in a way that is consistent with the effects of IGMF. 

The field strengths of IGMF inferred from this analysis are of the order of a femtogauss.  This result may open a new window on the particle physics in the early universe which can be responsible for generating the cosmological magnetic fields in a phase transition, inflationary reheating, or another out-of-equilibrium phenomenon\cite{Vachaspati:1991nm,Cornwall:1997ms,Kandus:2010nw,Kahniashvili:2010wm}.

\section{Protons and secondary photons from extragalactic sources}
 
As long as IGMF have strengths as low as few femtogauss~\cite{Ando:2010rb}, the protons from distant blazars travel essentially along straight lines and produce secondary photons in their interactions with photon backgrounds~\cite{Essey:2009zg,Essey:2009ju}.  For nearby sources, the primary gamma rays produced at the source dominate the ACT signals.  However, for more distant sources, the primary photons are filtered out by their interactions with EBL and the secondary photons can make up most of the observed signal~\cite{Essey:2009zg,Essey:2009ju}.

The secondary photons provide an excellent fit to the spectral shapes of gamma-ray signals from distant blazars~\cite{Essey:2009zg,Essey:2009ju}.  Such fits are essentially one-parameter fits that depend only on the overall power of the source emitted in cosmic rays.  The spectral slope of protons and the level of EBL do not have a strong effect on the spectrum of secondary photons.  However, for the same photon flux, the neutrino flux varies depending on the maximal energy $E_{\rm max}$ to which the protons are accelerated.  Indeed, there are two competing processes that generate secondary photons: $p\gamma_{EBL}\rightarrow p\pi^0 \rightarrow p\gamma\gamma $ and $p\gamma_{CMB}\rightarrow pe^+e^-$.  For smaller $E_{\rm max}$, a larger fraction of photons come from the hadronic channel, which is accompanied by production of neutrinos via $p\gamma_{EBL}\rightarrow n\pi^{\pm}$ followed by the decays of charged pions and the neutron.  
Neutrino observations can help determine this parameter~\cite{Essey:2009ju}. 

\begin{figure}[t!]
  \begin{center}
      \includegraphics[height=0.45 \textwidth,angle=270]{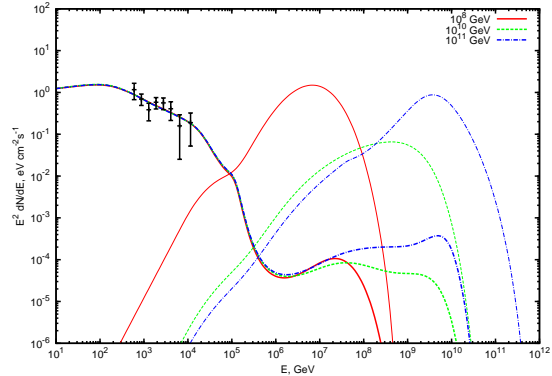}
\caption{Photon (low energy) and neutrino (high energy) spectra~\cite{Essey:2009ju} expected from an AGN at  $z=0.14$ (such as 1ES0229+200), normalized to HESS data points (shown)~\cite{Aharonian:2005gh}, for  $E_{\rm max}=10^8$GeV, $10^{10}$GeV, and $10^{11}$GeV shown by the solid, dashed, and dash-dotted lines, respectively.}
    \end{center}
  \label{fig:alpha2}
\end{figure}

The sources of secondary photons exhibit an unusual scaling law, which can be useful for population studies and for planning the future ACT.  The flux of protons scales with distance $d$ as $1/d^2$, but the rate of their interactions in the (optically thin) photon background is proportional to $d$.  Therefore, the flux of secondary photons scales as $1/d$, not $1/d^2$.   
This scaling with distance is the reason why the secondary photons, whose flux is proportional to $1/d$, become increasingly more important for distant sources, while the primary photon flux falls off more rapidly, as $\exp\{-d/\lambda\}/d^2$, where $\lambda $ is the attenuation length due to absorption. 

\section{Conclusions}

Based on the recent data, one can make several remarkable inferences about the ultrahigh-energy cosmic rays and magnetic fields inside the Milky Way and in the intergalactic space.  First, the energy dependent composition, with heavier nuclei at high energy, points to a non-negligible contribution from Galactic sources~\cite{Calvez:2010uh}.  Diffusion in turbulent Galactic magnetic field traps the nuclei more efficiently than protons, leading to an increase in the nuclear fraction up to the energy at which iron escapes ($\sim 30$~EeV).   At higher energies, the extragalactic protons should dominate the flux of UHECR, and their arrival directions should correlate with locations of the known sources.  Second, the recent measurement of intergalactic magnetic fields shows that the have {\em femtogauss} strengths~\cite{Ando:2010rb}, which means that high-energy protons generate secondary photons well aligned with the line of sight.  This allows one to study the cosmic ray output of AGN using atmospheric Cherenkov telescope observations of distant blazars.  Finally, if and when neutrino telescopes, such as IceCube~\cite{Halzen:2010yj} detect point sources, one can learn about the cosmic-ray sources and photon backgrounds by comparing the neutrino flux to the photon flux. Neutrino and gamma-ray observations can help distinguish the local Galactic sources from extragalactic sources of UHE nuclei~\cite{Murase:2010gj,Murase:2010va,Hooper:2010ze}. These inferences open exciting new opportunities for multi-messenger photon, charged-particle, and neutrino astronomy.  

The author thanks S.~Ando, J.~Beacom, A.~Calvez, W.~Essey, O.~Kalashev, and S.~Nagataki for collaboration and discussions of related topics. 
This work was supported by DOE Grant DE-FG03-91ER40662 and NASA ATFP Grant NNX08AL48G.






\end{document}